# Electronic decoupling and hole-doping of graphene nanoribbons on metal substrates by chloride intercalation

Amogh Kinikar[a], Thorsten G. Englmann[b], Marco Di Giovannantonio*[a,f], Nicolò Bassi [a], Feifei Xiang[a], Samuel Stolz[a,c], Roland Widmer[a], Gabriela Borin Barin[a], Elia Turco[a], Néstor Merino Díez[a], Kristjan Eimre[a,g], Andres Ortega Guerrero[a], Xinliang Feng[b,c], Oliver Gröning[a], Carlo A. Pignedoli*[a], Roman Fasel[a,e], and Pascal Ruffieux*[a]

[a]nanotech@surfaces Laboratory, Empa, Swiss Federal Laboratories for Materials Science and Technology, 8600 Dübendorf, Switzerland

[b]Center for Advancing Electronics Dresden & Faculty of Chemistry and Food Chemistry, TU Dresden, Dresden 01062, Germany

[c]Laboratory of Nanostructures at Surfaces, Institute of Physics, École Polytechnique Fédérale de Lausanne, 1015 Lausanne, Switzerland

[d]Max Planck Institute of Microstructure Physics, Weinberg 2, 06120 Halle, Germany

[e]Department of Chemistry, Biochemistry and Pharmaceutical Sciences, University of Bern, 3012 Bern, Switzerland

Present address:
[f]Istituto di Struttura della Materia – CNR (ISM-CNR), via Fosso del Cavaliere 100, Roma 00133, Italy
[g]EPFL, École polytechnique fédérale de Lausanne, NCCR MARVEL, Lausanne 1015, Switzerland

*Address correspondence to:
MDG: marco.digiovannantonio@artov.ism.cnr.it
CAP: carlo.pignedoli@empa.ch
PR: pascal.ruffieux@empa.ch

## ABSTRACT

Atomically precise graphene nanoribbons (GNRs) have a wide range of electronic properties that depend sensitively on their chemical structure. Several types of GNRs have been synthesized on metal surfaces through selective surface-catalyzed reactions. The resulting GNRs are adsorbed on the metal surface, which may lead to hybridization between the GNR orbitals and those of the substrate. This makes investigation of the intrinsic electronic properties of GNRs more difficult, and also rules out capacitive gating. Here we demonstrate the formation of a dielectric gold chloride adlayer that can intercalate underneath GNRs on the Au(111) surface. The intercalated gold chloride adlayer electronically decouples the GNRs from the metal and leads to a substantial hole doping of the GNRs. Our results introduce an easily accessible tool in the in situ characterization of GNRs grown on Au(111) that allows for exploration of their electronic properties in a heavily hole-doped regime.

## Introduction:

Carefully designed molecular precursors can be made to undergo selective surface-catalyzed reactions to yield atomically precise carbon nanomaterials[1–3]. The reactions that this so-called on-surface synthesis strategy relies on are best catalyzed by metallic surfaces, in particular Au(111). However, the carbon nanomaterials thus synthesized are found adsorbed on these catalytic surfaces, which can obscure the intrinsic electronic properties of the synthesized materials[4]. Moreover, the substrate also determines the Fermi energy of the adsorbed nanomaterials. This rules out any possibility for gating the nanomaterials without first electronically decoupling them from the metal.

To address these challenges, considerable efforts have been invested in establishing surface-catalyzed reaction protocols on insulating surfaces. However, successful nanomaterials synthesis on insulating substrates is rare and it requires the development of new surface-chemistries[5,6]. A promising strategy that combines the 'best-of-both-worlds' is to synthesize the desired nanomaterials on a metallic surface, and to subsequently intercalate a dielectric layer underneath the nanomaterials. This was previously achieved with Si and $I_2$ intercalation under graphene nanoribbons (GNRs) on Au(111)[7–10]. These, however, required annealing the substrates at temperatures greater than 200 °C, which led to undesirable chemical reactions[7]. Thus, there is an imperative need to find an intercalation strategy that works at lower temperatures where such undesirable chemical reactions are energetically prohibited.

Here we demonstrate the intercalation of gold chloride under 7-atom-wide armchair GNRs (7-AGNRs[11]) on Au(111), which occurs at room temperature. The formation of the gold chloride adlayer introduces an insulating layer between the metallic substrate and the GNRs and significantly hole-dopes the 7-AGNRs, transforming them from semiconductors into quasi-one-dimensional metals. We characterize the intercalated GNRs by scanning tunneling microscopy (STM), spectroscopy (STS), and non-contact atomic force microscopy (nc-AFM). Additional insights are obtained by photoelectron spectroscopy and density functional theory (DFT) calculations. Our findings provide a novel and practical strategy for intercalating dielectric layers underneath atomically precise carbon nanomaterials to tune their electronic properties.

## Results and Discussions

### I. Intercalation of gold chloride under GNRs

The prototypical atomically precise 7-AGNR is chosen as a model system to demonstrate the gold chloride intercalation strategy. First, the 7-AGNRs are synthesized on a clean Au(111) single-crystal surface, following established protocols[11]. After their synthesis, gold chloride intercalation is achieved *in situ* under ultrahigh vacuum (UHV) conditions, by the sublimation of AuCl onto the substrate held at room temperature (Fig. 1a, additional details in the methods section and in the Supporting Information S1). The STM image of the surface after the deposition of AuCl reveals a corrugated reconstruction with distinctive rows (Fig 1b). This reconstruction is clearly different from the herringbone reconstruction seen on clean Au(111) surfaces[12] and has not been observed when dosing $Cl_2$ onto Au(111) surfaces[13–18]. We present a detailed characterization of the structure and the electronic properties of this gold chloride adlayer in Section II, where we identify the stoichiometry of the adsorbed chloride adlayer to be $Au_2Cl_5$. The distinctive rows seen in the STM image (Fig. 1b) provide an easy means for identifying the formation of the $Au_2Cl_5$ adlayer. Moreover, we observe these to be contiguous under the GNRs (Fig. 1b, inset), indicating that the 7-AGNRs are adsorbed above the adlayer. This shows the successful intercalation of gold chloride under the 7-AGNRs at room temperature.

Interestingly, the STM image of the 7-AGNRs after the intercalation of the $Au_2Cl_5$ adlayer exhibits prominent contributions from the GNR's electronic states (Fig. 1b,c). Such prominent contributions of the electronic states have been observed only when the GNRs have been manipulated onto insulating NaCl islands on a metal surface[19], i.e. when the GNRs have been electronically decoupled from the metal. Additionally, the apparent height of the 7-AGNR as determined from the height profile in Fig. 1c is 2.8 Å, considerably higher than that of the GNR on Au(111) (~1.8 Å)[11]. These observations indicate that the intercalation of the $Au_2Cl_5$ adlayer electronically decouples the 7-AGNRs from the metal.

Considering that the halogens are very reactive, the present strategy could potentially chlorinate the GNRs. To confirm that the intercalation of the $Au_2Cl_5$ adlayer does not alter the chemical structure of the 7-AGNR, non-contact atomic force microscopy (nc-AFM) images[20–22] are acquired with a CO- (or possibly Cl-[20,23]) functionalized tip (Fig. 1d). Bond–resolved nc-AFM images clearly resolve the planar structure of the unaltered 7-AGNRs. Additionally, the distinctive rows of the adlayer also have a corrugation along their length. This corrugation is neither visible in the STM image (Fig. 1c) nor in the nc-AFM image (Fig. 1d) of the adsorbed GNR, suggesting that the interaction of the GNRs with the $Au_2Cl_5$ adlayer is weak.

While the GNRs appear structurally unaltered, a notable change in the electronic properties of the 7-AGNRs can be deduced by the appearance of molecular orbitals in the constant-height current map acquired at 10 mV bias (Fig. 1e). 7-AGNRs on Au(111) have a band gap of 2.36 eV with the conduction band (CB) minimum at 1.52 eV and valence band (VB) maximum at -0.84 eV[24]. For 7-AGNRs adsorbed on Au(111), a current map at 10 mV bias with a functionalized tip (as is the present case) would lead to a 'bond-resolved' image that resembles the chemical structure of the GNR[25]. However, after the intercalation of the $Au_2Cl_5$ adlayer the current-map instead is found to exhibit molecular orbital features. This indicates that some electronic states are present in the energy region between the applied bias voltage and the Fermi level. These intermediary states allow for resonant tunneling of electrons or holes from the STM tip to the underlying substrate. Such states are observed here at a bias voltage of only 10 mV, which implies that upon intercalation of the $Au_2Cl_5$ adlayer, the 7-AGNR has electronic states very close to the Fermi level, i.e. the 7-AGNR is now metallic. Since we do not observe any chemical changes to the 7-AGNR's structure, the insulator-to-metal transition can only arise because of charge doping.

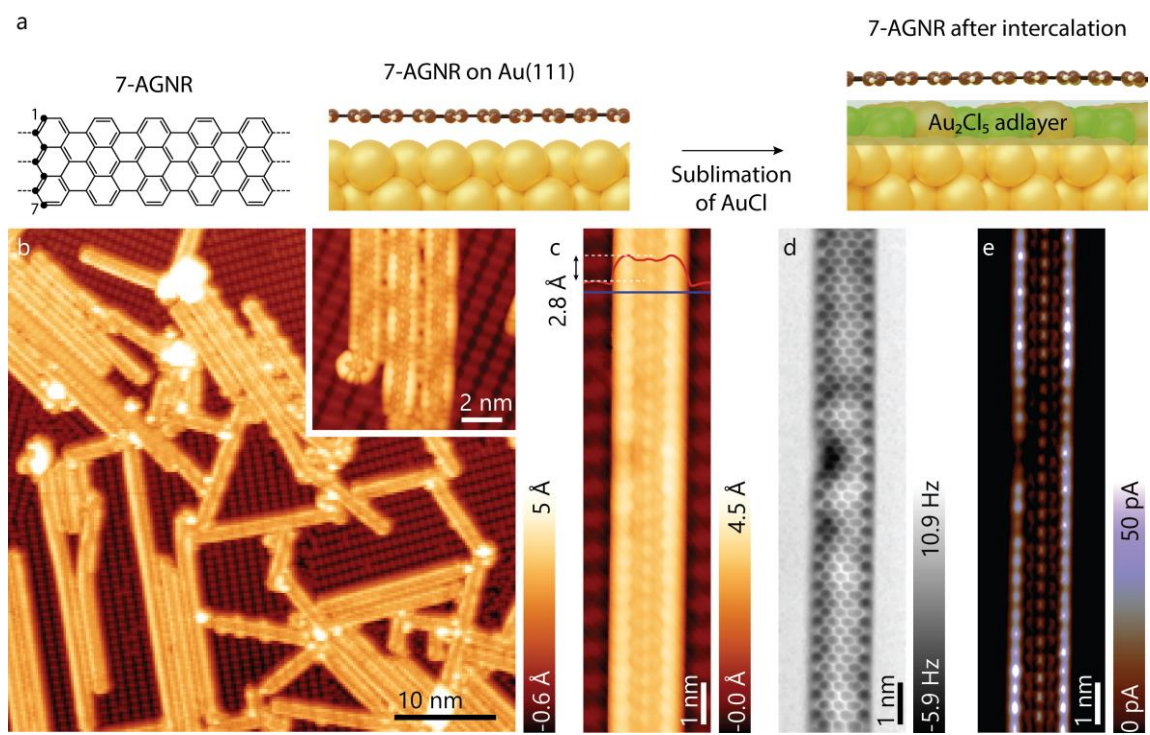

**Figure 1: Au$_2$Cl$_5$ adlayer intercalation under 7-AGNRs. a,** The chemical structure of the 7-AGNR is shown, marked with the 7 carbon atoms that define its width. The schematic illustrations shows the 7-AGNRs synthesized on the Au(111) surface. The in situ sublimation of AuCl leads to the intercalation of a Au$_2$Cl$_5$ adlayer underneath the 7-AGNRs. **b**, STM image showing the 7-AGNRs after the intercalation of gold chloride. Inset: Higher resolution image of the surface showing that the rows of the reconstruction are contiguous under the GNRs. (V = 1.5 V, I = 20 pA, inset: V = 0.05 V, I = 20 pA, both images are shown with the same color axis) **c**, STM image acquired over a 7-AGNR segment (V = 0.25 V, I = 20 pA). The height profile along the blue line is shown in the red curve overlaid on the STM image. **d**, The bond-resolved nc-AFM image shows the intact 7-AGNR with no significant corrugation in the 7-AGNR adsorption height (constant-height, feedback switched off (Δz = 0) on top of the GNR at 10 mV, 20 pA). **e**, Constant-height current map acquired at 10 mV exhibits electronic states near the Fermi level (feedback switched off on top of the GNR at 10 mV, 20 pA).

## II. Characterization of the gold chloride surface

In the previous section, we showed that the Au$_2$Cl$_5$ adlayer decouples the 7-AGNRs from the bulk Au substrate (evidenced by the adsorption height and prominent electronic state contributions in the STM image of the 7-AGNRs) and that the 7-AGNRs are electronically doped (evidenced by the induced metallicity). We now present a characterization of the Au$_2$Cl$_5$ adlayer that rationalizes the above observations.

Previous works[13–18] attempted to synthesize gold chloride on the Au(111) surface by dosing Cl$_2$ gas. Molecules of Cl$_2$ undergo dissociative chemisorption catalyzed by the Au surface. However, the adsorption of Cl poisons the catalytic activity of the gold, and thus full monolayer coverage of gold chloride cannot be achieved by this means. To circumvent this, we achieve the $Au_2Cl_5$ adlayer by sublimating AuCl onto the Au(111) surface. The resulting surface chloride, with its well-ordered corrugated reconstruction, is at saturation. Sublimation of additional AuCl does not lead to a denser packing of the atoms in the adlayer nor does it lead to the formation of additional layers on top of it – the growth of the $Au_2Cl_5$ adlayer is thus self-limiting. Upon obtaining atomically resolved STM images of this saturated chloride layer, we find that the unit cell of this surface structure is comprised of seven atoms (Fig. 2a). As we will show in the following discussion, these seven atoms correspond to an $AuCl_2$ dimer[13,14,16] and one chemisorbed Cl atom leading to a stoichiometric ratio of $Au_2Cl_5$. These atoms are labeled in Fig. 2c with colored circles. High-resolution X-ray photoelectron spectroscopy (XPS) measurements of the Cl 2p core level reveal two spin-orbit doublets corresponding to the two different chlorine species (Fig. 2f). The XPS spectra can be fitted with excellent agreement with two doublets in a 4:1 ratio (see Methods for further details on the XPS measurements). UPS measurements reveal that the $Au_2Cl_5$ adlayer has a work function of 6.15 eV (Fig. 2g), which is 0.82 eV higher than the work function of the clean Au(111) surface (= 5.33 eV[26]) and is consistent with chlorine's strong electronegativity[27]. Considering vacuum level alignment between the $Au_2Cl_5$ adlayer and the weakly interacting 7-AGNRs, significant hole-doping is thus to be expected in the adsorbed GNRs.

Ab initio calculations of the $Au_2Cl_5$ adlayer on Au(111) further establish its structure. DFT calculations reveal that the $Au_2Cl_5$ adlayer has a non-orthorhombic unit cell with $\begin{bmatrix} 4 & -1 \\ -1 & 3 \end{bmatrix}$ registry with the lattice vectors of the underlying Au(111) surface. The optimized $Au_2Cl_5$ adlayer structure along with its unit cell expressed in terms of the Au(111) surface lattice vectors is shown in Fig. 2c,d (see Supporting Information S2). Simulated STM images (Fig. 2b) on this geometry are in good agreement with the experiment (Fig. 2a), providing further support to the proposed structure.

Moreover, as a metal halide, the $Au_2Cl_5$ adlayer is expected to be an insulator. Scanning tunneling spectroscopy (STS) measurements on the $Au_2Cl_5$ adlayer show that it indeed has a large bandgap of 3.4 eV (Fig. 2e). Thus, the formation of the $Au_2Cl_5$ adlayer presents a straightforward way to transform the catalytically active, metallic gold surface into an insulator. The intercalation of an insulating layer adds a tunneling barrier between the electrons in gold and the electrons in the GNRs. This exponentially weakens the hybridization between the electronic states of the two systems, thereby electronically decoupling them.

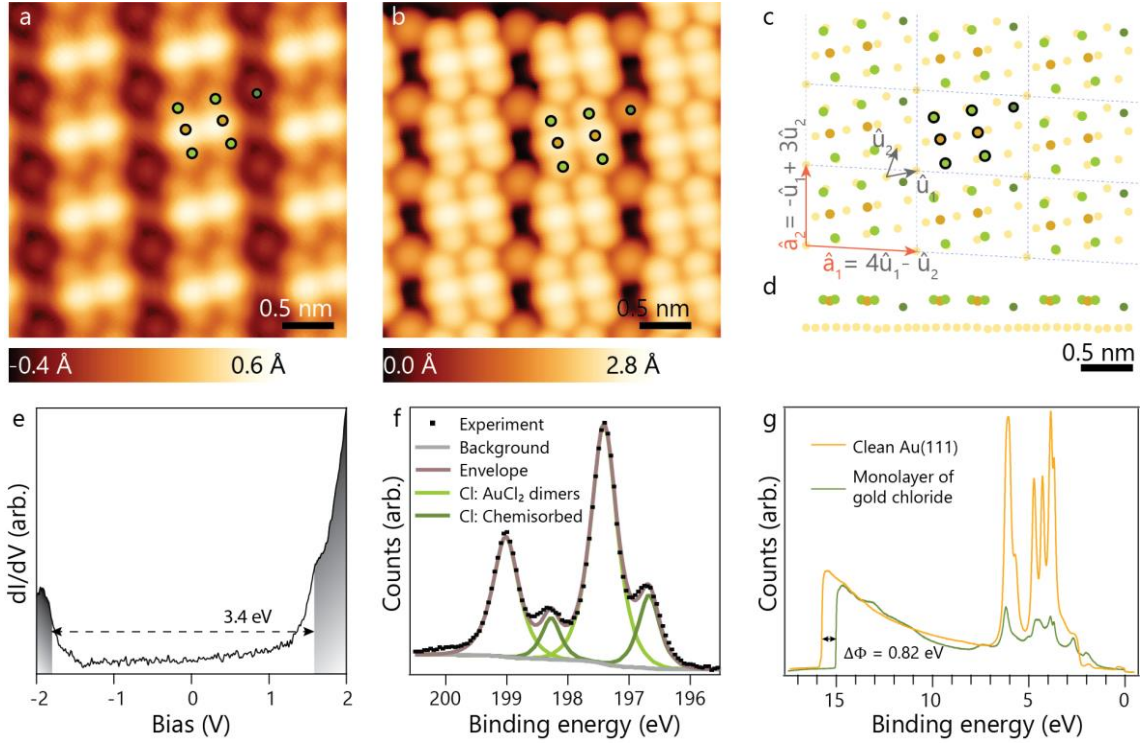

**Figure 2: Characterization of the gold chloride layer. a**, Atomically resolved, experimental STM image reveals seven atoms in the unit cell of the $Au_2Cl_5$ adlayer (V = 10 mV, I = 100 pA). **b**, Simulated STM image on the DFT-optimized geometry of the $Au_2Cl_5$ adlayer on Au(111) (V = 50 mV). **c**, Top view of the DFT-optimized geometry of the $Au_2Cl_5$ adlayer represented along with the first layer of Au(111) atoms. The Au adatoms are represented in dark yellow and the Au atoms in the substrate are shown in light yellow. The chemisorbed Cl is shown in dark green, the Cl in the $AuCl_2$ in light green. The outlined atoms are those indicated on the STM images in **a** and **b**. The $Au_2Cl_5$ layer lattice vectors $\hat{a}_1$ and $\hat{a}_2$ are shown in terms of the surface lattice vectors of Au(111) $\hat{u}_1$ and $\hat{u}_2$: $\hat{a}_1 = 4\hat{u}_1 - \hat{u}_2$ and $\hat{a}_2 = -\hat{u}_1 + 3\hat{u}_2$. **d**, Side view of the DFT-optimized geometry of the $Au_2Cl_5$ adlayer, highlighting the higher adsorption height of the $AuCl_2$ dimers (2.6 Å) compared to the chemisorbed Cl (1.9 Å) (the scale bar for **c** and **d** is shown below **d**). **e**, The STS measurements show that the band gap of the $Au_2Cl_5$ adlayer is ~3.4 eV (feedback switched off at 1 V, 20 pA. Excitation voltage = 14 mV). **f**, High-resolution XPS spectrum acquired at the Cl 2p core level with 425 eV synchrotron radiation. The spectrum can be deconvoluted into two Cl species with Cl $2p_{3/2}$ peaks at 197.4 eV for the Cl in $AuCl_2$ dimers and 196.7 eV for the chemisorbed Cl. The corresponding experimental peak ratios agree with a 4:1 ratio of the two Cl species. See Methods for experimental and fitting details). **g**, UPS spectra of clean Au(111) and after the formation of the $Au_2Cl_5$ adlayer. The work function Φ of the $Au_2Cl_5$ adlayer covered surface is 6.15 eV, 0.82 eV higher than the work function of the clean Au(111) surface (5.33 eV). See Methods for experimental details.

### III.    Characterization of short 7-AGNR segments adsorbed on $Au_2Cl_5$ adlayer

Having characterized the $Au_2Cl_5$ adlayer, we proceed to investigate the hole-doping resulting from the increase of the work function upon the formation of the adlayer. To characterize the induced doping of the 7-AGNRs, we examine short 7-AGNR segments, namely teranthene (**2**) and hexanthene (**3**), which have well-separated energy levels due to size confinement. These energy levels can be clearly identified using STS measurements before and after intercalation, allowing for a quantification of the energy shift of each level. We obtain **2** and **3** by annealing 10-bromo-9,9':10',9''-teranthracene (**1**) to 350 °C on the Au(111) surface (Fig. 3a) following the method reported by Borin Barin et al[28]. The $Au_2Cl_5$ adlayer is then formed by sublimation deposition of AuCl. STM images of the resulting surface show a well-ordered $Au_2Cl_5$ adlayer underneath the short 7-AGNR segments, with individual molecules dispersed across the crystal terrace (Fig. 3b).

STS measurements on **2** and **3** (Fig 3c,d) reveal sharp peaks deriving from resonant tunneling through molecular levels. Differential conductance (dI/dV) maps acquired at the bias voltage of the peaks in the STS spectra provide a spatial map of the corresponding molecular orbitals. These are in excellent agreement with DFT simulations of the gas-phase orbitals for both molecules and allow for the unambiguous identification of the observed molecular levels.

Unsurprisingly, the energy at which these molecular levels occur is significantly different in gas-phase (by DFT, for a charge-neutral system) and on the $Au_2Cl_5$ adlayer (as determined experimentally). In case of **2**, the highest occupied molecular orbital (HOMO) in gas-phase is fully depopulated when **2** is adsorbed on $Au_2Cl_5$ adlayer and is observed at 0.49 eV. Depopulation of the gas-phase HOMO is also observed in case of **3** when adsorbed on the $Au_2Cl_5$ adlayer, and is observed at 1 eV. Table 1 summarizes the results of our measurements and simulations. It also includes the previously published spectroscopic characterization of **2** and **3** on Au(111). With respect to the experimentally observed values of the orbitals when **2** and **3** are adsorbed on Au(111), the intercalation of the $Au_2Cl_5$ adlayer increases the molecular orbital energies by 0.6 – 0.9 eV as a consequence of substantial hole-doping of the adsorbates, providing clear evidence for charge transfer (see Supporting Information S3).

|   | Gas-Phase Orbital Index | Gas-Phase DFT (eV) | On Au(111)[28] (eV) | on $Au_2Cl_5$ adlayer (eV) | Shift in Energy (eV) |
|---|---|---|---|---|---|
|   |   | **A** | **B** | **C** | **C - B** |
| **2** | HOMO | -0.17 | -0.10 | 0.49 | 0.59 |
|   | LUMO | 0.17 | 0.20 | 0.88 | 0.68 |
| **3** | HOMO -1 | -1.01 | -1.10 | -0.62 | 0.48 |
|   | HOMO | -0.02 | 0.10 | 1.00 | 0.90 |
|   | LUMO | 0.02 | 0.44 | 1.34 | 0.90 |

**Table 1: Energies of the orbitals of 2 and 3 in gas phase (by DFT), and experimentally determined by STS on Au(111) and on the $Au_2Cl_5$ layer/Au(111).** The energies of the indicated orbitals are shown: **A** is the energy of the orbitals in gas phase determined by DFT. **B** is the energy

of the orbitals as measured on an Au(111)[28] . **C** is the measured after the intercalation of $Au_2Cl_5$ adlayer under the molecules shown in Fig. 3.

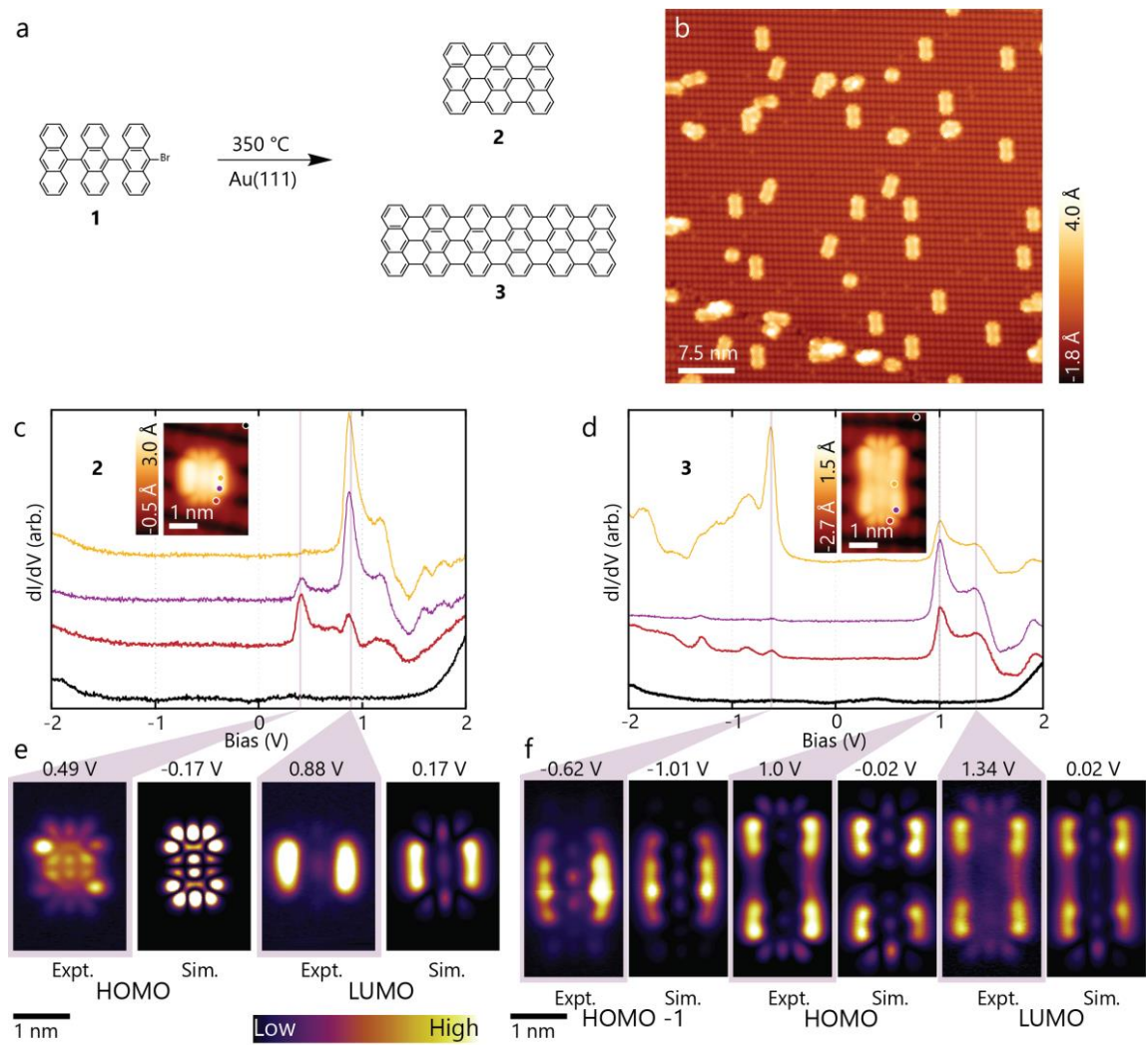

**Figure 3: STS characterization of 2 and 3 on the $Au_2Cl_5$ adlayer. a**, Schematic showing the synthesis of teranthene (**2**) and hexanthene (**3**) on Au(111). **b**, STM image acquired after the intercalation of $Au_2Cl_5$ adlayer. **c**-**d**, STS spectra acquired on **2** and **3**. The acquisition points are indicated in the STM images shown in the insets (STM images of **2** (**a**, V = 2 V, I = 100 pA) and **3** (**b**, V = 2 V, I = 150 pA)). The spectra are color-coded according to the positions over which they were obtained and vertically offset for clarity. (feedback switched off over the indicated positions at 2 V, 100 pA (**a**) and 2 V, 150 pA (**b**), peak-to-peak excitation voltage: 20 mV). **e**-**f**, Constant-height dI/dV maps acquired at the bias values of the prominent STS peaks over the two molecules (feedback switched off on top of the molecules at the indicated bias values and I = 100 pA for (**c**) and I = 150 pA for (**d**), peak-to-peak excitation: 20 mV).The dI/dV maps are in good agreement with the gas-phase orbitals of **2** and **3** obtained by DFT that are shown adjacent to the experimental maps and labeled with the corresponding orbital and its gas phase energy.

## IV. Hole-doping induced metallicity in 7-AGNR

Capacitive gating is a crucial tool for characterizing the electronic properties of materials, as it can modulate the number of charge carriers within a system. Notably, in systems characterized by strong electron-electron interactions, doping can trigger quantum phase transitions. In the case of the $Au_2Cl_5$ adlayer, its minimal thickness precludes the effective utilization of capacitive gating of the adsorbed GNRs. Nevertheless, a noteworthy observation is the significant hole-doping (**C-B**, Table 1) upon intercalation of the $Au_2Cl_5$ adlayer. While this doping level remains fixed and lacks tunability, it presents an intriguing avenue for investigating the electronic characteristics of the adsorbed system when it is charged.

A preliminary finding highlighting the potential of intercalation-induced doping is the emergence of electronic states near the Fermi level for longer 7-AGNRs adsorbed on the $Au_2Cl_5$ adlayer (Fig. 1e). While the presence of such near Fermi level states suggests a metallic nature of the system, it is essential that these states are delocalized, which means that the Fermi level must intersect a dispersive band. In finite-size systems, dispersive electrons display particle-in-a-box behavior as they scatter from the system's ends[24]. The resulting modulation in the local density of states (LDOS) can be subjected to a Fourier transform to directly access the dispersion relation in reciprocal space. The achievable resolution $\delta q = 2\pi/L$ in reciprocal ($q$)-space is determined by the experimentally probed length $L$ of the system under investigation, defined by the number of data acquisition points and their spacing $(L = N\delta x)$. High resolution in reciprocal space thus requires data acquisition on a sufficiently long system in real space. Moreover, as the wavelength of the wave function is twice the wavelength of its probability density, the wavevector $k = q/2$. We thus select a sufficiently long 7-AGNR (Fig. 4a) and acquire STS spectra along its edge. The STS spectra, when visualized with respect to the acquisition position, distinctly reveal the particle-in-a-box states (Fig. 4b), and a line-by-line fast Fourier transform exposes the dispersive valence band (VB) in reciprocal space (Fig. 4c). The dispersion of the VB in the 7-AGNR near $k = 0$ can be accurately described by a nearly free-electron gas model with an effective mass $m^*$, with energy dispersion given by[29,30]:

$$E(k) = E_{k=0} + \frac{\hbar^2}{2m^*}k^2$$

By fitting the observed dispersion $E(k)$ (Fig. 4c) to a parabola as defined above, the effective mass is found to be $m^* = 0.32 \pm 0.02$ $m_e$. This is identical to the effective mass obtained for the 7-AGNR VB on NaCl islands on Au(111)[19]. The valence band maximum ($E_{k=0}$) is found at +187 meV, demonstrating that the intercalation of the $Au_2Cl_5$ adlayer indeed induces hole-doping in the 7-AGNRs (see Supporting Information S4 for a real-space comparison between the GNR and particle-in-a-box states). We note that the second valence band (VB-1) of the 7-AGNR is not detected in the data shown in Fig. 4, as probing it would require approaching the tip closer to the GNR which is not feasible in this case due to the weak adsorption of the GNRs on the $Au_2Cl_5$ adlayer[31].

One-dimensional systems cannot be truly metallic, because electron-electron and electron-phonon interactions necessarily lead to the opening of a gap at the Fermi level[32]. However, for the finite-length GNR under study, the dominant energy scale is the level spacing (~10 meV) which is much larger than the thermal energy scale of ~0.4 meV. This implies that it behaves as a 0-D object rather than a 1-D one. However, by increasing the length of the system under study, the level-spacing can be made infinitesimally small. For sufficiently large systems, for example several micrometer long carbon nanotubes, the level spacing would be smaller than the thermal energy scale, manifesting correlated electron physics expected of 1D metals. Presently, the on-surface synthesis of several micrometer long GNRs has not been achieved. The level spacing could also be decreased by increasing the effective mass of the charge carriers; however, this would involve designing and synthesizing a new atomically precise GNR and is therefore beyond the scope of the present work.

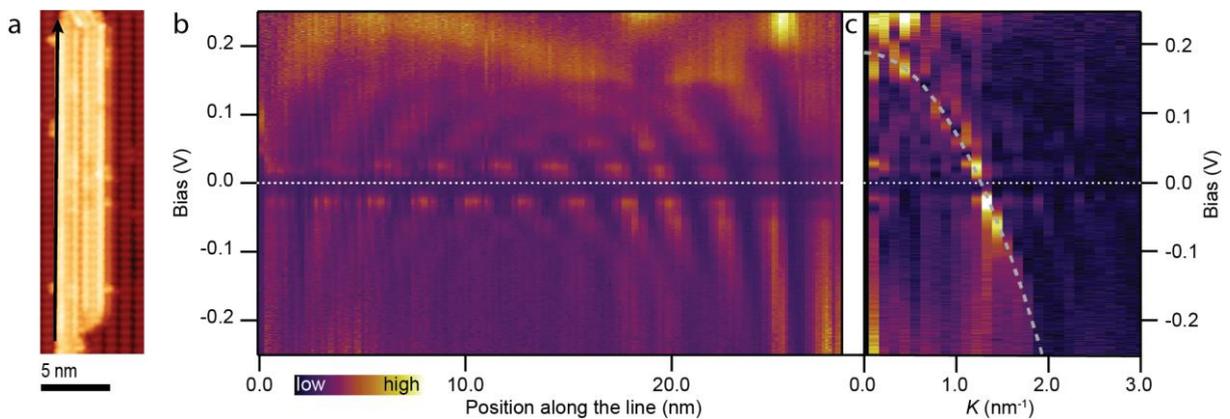

**Figure 4: Dispersion of the 7-AGNR valence band. a**, STM image of two parallel 7-AGNRs on the $Au_2Cl_5$ layer (V = 1.0 V, I = 10 pA). STS spectra have been acquired along the black line at the edge of the left, 28 nm long 7-AGNR. **b**, STS spectra plotted versus acquisition position, revealing particle-in-a-box states (dI/dV point spectra were acquired on 175 equidistant points with spacing of δx= 0.160 nm. The feedback was switched off on each point at 0.25 V, 200 pA. Excitation voltage = 1.414 mV). **c**, Fourier transform of the dI/dV versus position map in b with the fitted parabolic dispersion (grey dashed line).

## Conclusion

The atomically precise on-surface synthesis of GNRs is, with a few but promising exceptions, confined to metal surfaces. Here we have demonstrated that the Au(111) surface can be transformed into a dielectric layer by the deposition of gold chloride. The resulting gold chloride adlayer, which has a stoichiometry of $Au_2Cl_5$, can intercalate under pre-synthesized GNRs and electronically decouple them from the metal substrate. Moreover, the $Au_2Cl_5$ adlayer hole-dopes the intrinsically semiconducting 7-AGNRs into a metallic phase. The substantial charge transfer induced by $Au_2Cl_5$ intercalation may be used to dope GNRs with suitable bandgaps and bandwidths into exotic states manifesting many-body physics. The possibility to study the large variety of carbon nanomaterials fabricated via on-surface synthesis on Au(111) in a decoupled and hole-doped regime opens up new and exciting prospects.

## Methods

### Intercalation of the Au$_2$Cl$_5$ adlayer

The experiments were performed under UHV conditions (base pressure ~5×10$^{-10}$ mbar). Au(111) substrates (MaTeck GmbH) were cleaned by repeated cycles of Ar$^+$-ion sputtering (1 keV) and annealing (400-430 °C). GNRs were grown as per previous reports[11,28]. After GNR growth, the surface was cooled down to room temperature before the deposition of AuCl. AuCl (gold (I) chloride) was obtained from Sigma Aldrich. The salt was added to a quartz crucible and transferred into a custom evaporator in the UHV system (see Supporting Information S1). The crucible was then heated to 70-100 °C and allowed to stabilize for 15 min. The Au(111) surface was then exposed to the AuCl, and the Au$_2$Cl$_5$ adlayer was obtained after 90-110 min of deposition. We found that a brief annealing to temperatures less than 100 °C after the formation of the adlayer led to more stable imaging conditions (see Supporting Information S5). STM images acquired at different stages of this process are shown in the Supporting Information S6.

### XPS Measurements

The XPS experiments were performed on the X03DA (PEARL) beamline at the Swiss Light Source, Paul Scherrer Institut, Villigen, Switzerland. Linearly polarized light with photon energy of 425 eV was used to acquire XPS spectra. The spectra were obtained in normal emission geometry, using a Scienta R4000 hemispherical electron analyzer equipped with a multichannel plate (MCP) detector. The spectra were recorded in swept mode, with a step size of 50 meV and pass energy of 20 eV. The sample was cooled using liquid Helium, and the surface temperature was ~ 30 K. The Cl 2p spectra were fitted in CasaXPS with pseudo-Voigt line shapes (GL(80)) after subtraction of a Shirley background. The area of the 2p$_{3/2}$ peak corresponding to chemisorbed Cl was constrained to be one-fourth of the area of the 2p$_{3/2}$ peak obtained for the Cl in AuCl$_2$ dimers to match the stoichiometry of the Au$_2$Cl$_5$ adlayer. If the area constraints were removed, the ratio of the areas under the peaks was still reasonable at 1 : 3.6. Additionally, in all cases 2p$_{1/2}$ peak areas were constrained to be one half of the corresponding 2p$_{3/2}$ peak areas.

### UPS Measurements

The UPS measurements were performed in a combined LT-SPM & Photoemission system from Scienta Omicron. A focused Helium (I) gas discharge lamp (HIS 14 Focus GmbH) was used as the UV light source (Photon energy = 21.21 eV). The pressure inside the analysis chamber during the operation of the lamp was 3.5×10$^{-9}$ mbar. Spectra were acquired by a Scienta R3000 display analyzer in normal emission, in steps of 5 meV and with a pass energy of 20 eV. The Au$_2$Cl$_5$ adlayer on Au(111) was synthesized and checked by STM, the sample was subsequently transferred to the attached XPS-UPS analysis chamber. The sample was cooled down to ~ 30 K using liquid helium and biased at 9.53 V using a floating DC power supply for the work function measurements.

**STM, STS and nc-AFM Measurements**

STM/nc-AFM measurements were performed in two low-temperature scanning tunneling microscopes (Scienta Omicron) operated at 4.7 K. Both systems have similar base pressure (< $10^{-10}$ mbar) and sample preparation facilities. The first system is equipped with a SRS 830 lock-in amplifier. The second system is equipped with a tungsten tip placed on a qPlus® tuning fork sensor[33] and HF2Li PLL by Zurich Instruments. The differential conductance was measured using the lock-in technique, the peak-to-peak excitation voltages used are provided in the figure captions. The Fast Fourier Transform of the STS versus position data was performed in Igor Pro (Wavemetrics).

For the nc-AFM measurements the tip was functionalized with a single CO molecule at the tip apex picked up from the previously CO-dosed surface[34]. However, identifying the CO on the highly corrugated $Au_2Cl_5$ layer was challenging. The CO molecule was picked up from the $Au_2Cl_5$ layer on Au(111) by scanning the surface at -10 mV and 120-200 pA, however we cannot rule out if the tip was accidentally functionalized with Cl. The sensor was driven at its resonance frequency (24744 Hz, quality factor 20.6k) with a constant amplitude < 100 pm. The frequency shift from the resonance of the tuning fork was recorded in constant-height mode and measured using the PLL. The Δz is positive when the tip-surface distance is increased with respect to the STM set point and negative when the tip-surface distance is decreased. Δz is set to zero when the feedback loop is switched off.

All data was acquired using the Omicron Matrix electronics and processed using the MatrixFile Reader (byte-physics) XOP in Igor Pro[35].

**Computational Details**

All DFT calculations were performed with AiiDAlab platform[36] based on AiiDA[37] workflows for the CP2K code[38].

To find the equilibrium geometry of the $Au_2Cl_5$ layer structure, starting geometries were guessed from previous works on Cl adsorption on Au(111)[14,16], the atomic-resolution STM images and the high-resolution XPS measurements. An orthorhombic guess for the gold chloride structure did not yield a geometry that agreed with experiments. However, starting from a non-orthorhombic guess led to a structure that was in excellent agreement with our experiments. To simulate the STM images, an island of the optimized $Au_2Cl_5$ adlayer was placed on a rectangular Au(111) slab and the structure was then further optimized. STM simulations were obtained on this relaxed slab (see supporting information S2). Additionally, molecules of **2** and **3** were placed on top of this $Au_2Cl_5$ adlayer on orthorhombic Au(111) slab and their geometry optimized. This geometry was subsequently used for calculating the gas-phase orbitals shown in Fig. 4.

All structures were modeled within the repeated slab scheme. The simulation cell consisted of 4 atomic layers of Au along the [111] direction. For the $Au_2Cl_5$ adlayer, 6 atomic layers of Au were used, revealing that the 4-Au layers were sufficient for finding the optimized geometry. Consequently, to optimize the geometry of an $Au_2Cl_5$ island on a rectangular Au-slab, a slab consisting of 4 atomic layers (excluding the $Au_2Cl_5$ adlayer) was used(Supporting Information S2). A layer of hydrogen atoms was used to passivate one side of the slab to suppress the Au(111) surface state. 40 Å of vacuum was included in the simulation cell to decouple the system from its periodic replicas in the direction perpendicular to the surface. The electronic states were expanded with a TZV2P Gaussian basis set[39] for C, Cl and H species, and a DZVP basis set for Au species. A cutoff of 600 Ry was used for the plane-wave basis set. Norm-conserving Goedecker-Teter-Hutter pseudopotentials[40] were used to represent the frozen core electrons of the atoms. We used the PBE parameterization for the generalized gradient approximation of the exchange-correlation functional.[41] To account for van der Waals interactions, we used the D3 scheme proposed by Grimme.[42] The gold surface was modeled using a supercell, with its size ranging from 33.5 × 31.0 Å$^2$ (corresponding to 792 Au atoms) to 58.9 × 61.1 Å$^2$ (1920 Au atoms). To obtain the equilibrium geometries, we kept the atomic positions of the bottom two layers of the slab fixed to the ideal bulk positions, and all other atoms were relaxed until forces were lower than 0.005 eV/Å. STM images were simulated within the Tersoff-Hamann approximation[43] based on the Kohn-Sham orbitals of the slab/adsorbate systems. The orbitals were extrapolated to the vacuum region in order to correct the wrong decay of the charge density due to the localized basis set[40].

## Acknowledgements

This work was supported by the Swiss National Science Foundation (No. 200020_182015), the NCCR MARVEL funded by the Swiss National Science Foundation (No. 205602), the Werner Siemens-Stiftung (CarboQuant), and the Max Planck Society. Computational support from the Swiss Supercomputing Center (CSCS) under project ID s1141 is gratefully acknowledged. We acknowledge PRACE for awarding access to the Fenix Infrastructure resources at CSCS, which are partially funded from the European Union's Horizon 2020 research and innovation program through the ICEI project under the grant agreement No. 800858. S.S. acknowledges funding from the Swiss National Science Foundation under the project numbers 159690 and 195133. F.X. thanks the Walter-Benjamin Fellowship from Deutsche Forschungsgemeinschaft (DFG) and the SNSF Postdoctoral Fellowship. Technical support from Lukas Rotach is gratefully acknowledged. The XPS experiments were performed on the X03DA (PEARL) beamline at the Swiss Light Source, Paul Scherrer Institut, Villigen, Switzerland. We thank the beamline manager Dr. Matthias Muntwiler (PSI) for his technical support.

# Supporting Information

## Electronic decoupling and hole-doping of graphene nanoribbons on metal substrates by chloride intercalation

Amogh Kinikar[a], Thorsten G. Englmann[b], Marco Di Giovannantonio[a,f], Nicolò Bassi [a], Feifei Xiang[a], Samuel Stolz[a,c], Roland Widmer[a], Gabriela Borin Barin[a], Elia Turco[a], Néstor Merino Díez[a], Kristjan Eimre[a,g], Andres Ortega Guerrero[a], Xinliang Feng[b,c], Oliver Gröning[a], Carlo A. Pignedoli[a], Roman Fasel[a,e], and Pascal Ruffieux[a]

[a]nanotech@surfaces Laboratory, Empa, Swiss Federal Laboratories for Materials Science and Technology, 8600 Dübendorf, Switzerland
[b]Center for Advancing Electronics Dresden & Faculty of Chemistry and Food Chemistry, TU Dresden, Dresden 01062, Germany
[c]Laboratory of Nanostructures at Surfaces, Institute of Physics, École Polytechnique Fédérale de Lausanne, 1015 Lausanne, Switzerland
[d]Max Planck Institute of Microstructure Physics, Weinberg 2, 06120 Halle, Germany
[e]Department of Chemistry, Biochemistry and Pharmaceutical Sciences, University of Bern, 3012 Bern, Switzerland

Present address:
[f]Istituto di Struttura della Materia – CNR (ISM-CNR), via Fosso del Cavaliere 100, Roma 00133, Italy
[g]EPFL, École polytechnique fédérale de Lausanne, NCCR MARVEL, Lausanne 1015, Switzerland

## S1: Additional details on the sublimation of AuCl

Gold (I) chloride (AuCl, 99.9 % trace metal purity) was obtained from Sigma-Aldrich. AuCl is a metastable salt and in presence of water, it forms gold (III) chloride ($Au_2Cl_6$) and Au. The acquired AuCl was transferred into a diaphragm-sealed vial. The vial was evacuated and then filled with dry argon gas. Despite this, over a period of two years, the powder became wet due to deliquescence. AuCl itself is insoluble in water, but $Au_2Cl_6$ is highly hygroscopic and can form an aqueous solution by absorbing water from air. The presence of water thus makes the formation of $Au_2Cl_6$ from AuCl a chain reaction as any $Au_2Cl_6$ would solvate, and the water would trigger the formation of more $Au_2Cl_6$. Both gold chlorides decompose to release chlorine gas when heated under ambient conditions, AuCl from 170 °C and $Au_2Cl_6$ at 250 °C[1,2].

Based on the observed deliquescence, the gold chloride used for sublimation here was a mixture of AuCl and $Au_2Cl_6$ (=$AuCl_x$). Before UHV sublimation, the wet $AuCl_x$ was transferred to quartz crucibles. The salt was then dried by heating the crucible in a custom-built HV chamber (base pressure ~$5\times10^{-8}$ mbar) where the crucibles can be annealed and their sublimation rate monitored by a quartz crystal microbalance (QCM) as a function of temperature. Two to four cycles of heating to 100 °C were required to dry the $AuCl_x$. During the first drying process the pressure inside the chamber spiked up to $10^{-3}$ mbar. The annealing process was repeated until no pressure spikes were observed. The observed outgassing correlated with how long the $AuCl_x$ in the crucible had been exposed to air. After drying, the crucible was immediately transferred from the HV chamber into the load-lock chamber (base pressure ~$5\times10^{-8}$ mbar) attached to the UHV chambers of the instruments through air.

Inside the UHV chamber, the crucibles were heated in a custom-built evaporator and the sublimation rate was monitored using a QCM. The first heating cycle in the UHV chamber resulted in outgassing of ~$1\times10^{-8}$ mbar. However, the maximum pressure in each subsequent cycle decreased, and within 5 cycles the crucible could be heated to reach the desired sublimation rate with the pressure in the chamber not exceeding $3\times10^{-9}$ mbar. The temperature of the crucible was then increased until the flux measured by the QCM was 1 Å/min. The temperature required to reach this rate was 70-100 °C. The temperature range reflects different crucibles, which contained different amounts of gold chloride and would thus reach the same flux at different temperatures. Under UHV conditions, it is likely that the Cl released by thermal decomposition of the gold chlorides escapes as atomic Cl, rather than as $Cl_2$. Previous measurements show that Au is also deposited along with Cl in a 1: 4 Au : Cl ratio[3] . We speculate that the Au is deposited onto the surface as molecules of AuCl or less plausibly $Au_2Cl_6$ Since we only observe chemisorbed Cl at low coverages (below 0.33 ML) of Cl on Au(111), the molecules of AuCl and $Au_2Cl_6$ disassociate into Au atoms and Cl atoms when adsorbed on the Au(111) surface, and the resulting Au atoms migrate to the step edges, incorporating into the bulk.

## S2: DFT optimization of the Au$_2$Cl$_5$ layer

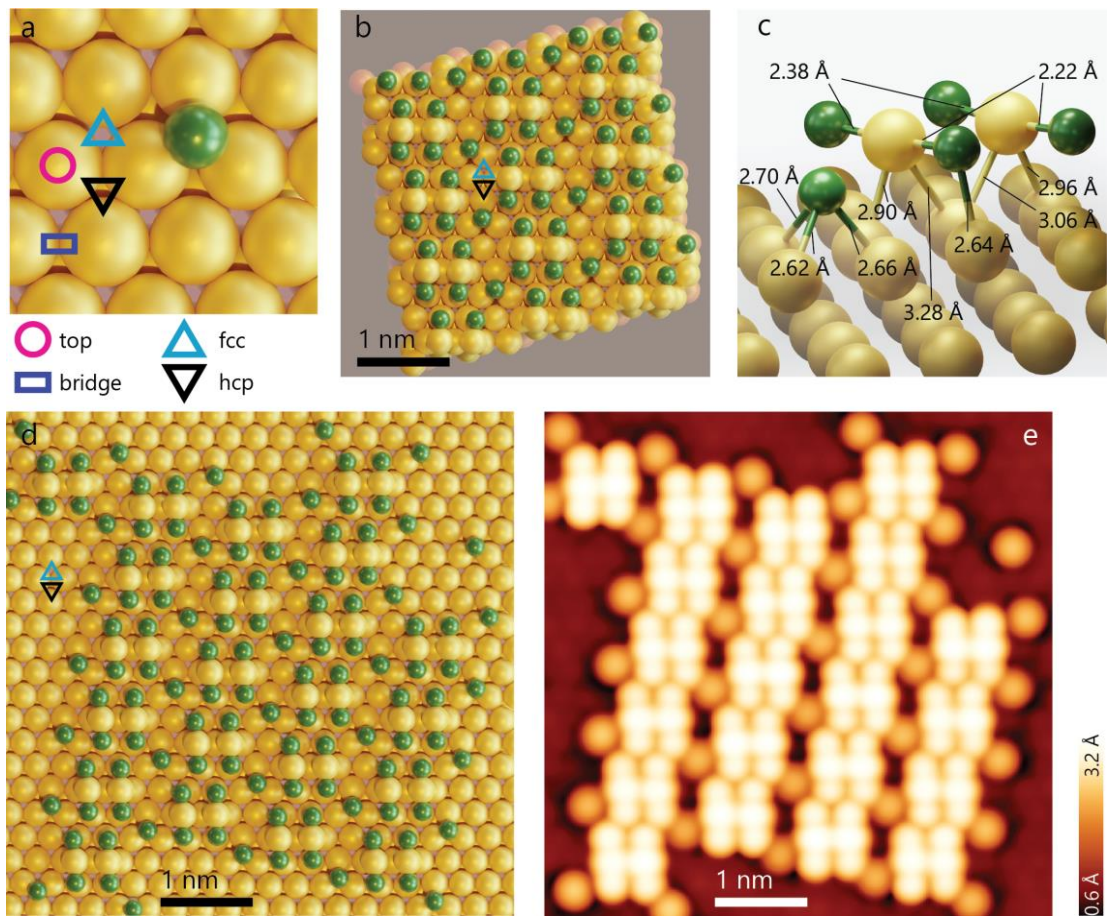

**Figure S1: DFT optimization of the Au$_2$Cl$_5$ layer structure. a**, The Au(111) surface has four high-symmetry adsorption sites: on top of an Au atom (top, pink circle), bridging two Au atoms (bridge, blue rectangle), an *fcc*-type hollow site (*fcc*, cyan triangle) and an hcp-type hollow site (*hcp*, black triangle). **b**, Rendered image of the DFT-optimized geometry of the Au$_2$Cl$_5$ adlayer. The Au adatoms are adsorbed on the bridge sites and the Cl atoms forming part of the AuCl$_2$ dimer are in approximately the top sites. The Cl atom adsorbed closest to the surface is in the *fcc* hollow site. **c**, The bond-lengths of the various atoms of the optimized geometry are indicated, only one Au$_2$Cl$_5$ cluster is shown for clarity. The Cl in the AuCl$_2$ dimers are 2.64 Å away from the nearest Au atom in the Au surface, and are thus more strongly bound to the Au adatom than to the surface. **d**, Optimized geometry of an Au$_2$Cl$_5$ adlayer island on a rectangular Au(111) slab. **e**, Simulated STM image over the optimized geometry in **d** obtained at V = -50 mV. This data was also shown in Fig. 2b in the main text.

## S3: Shift in the orbital energies

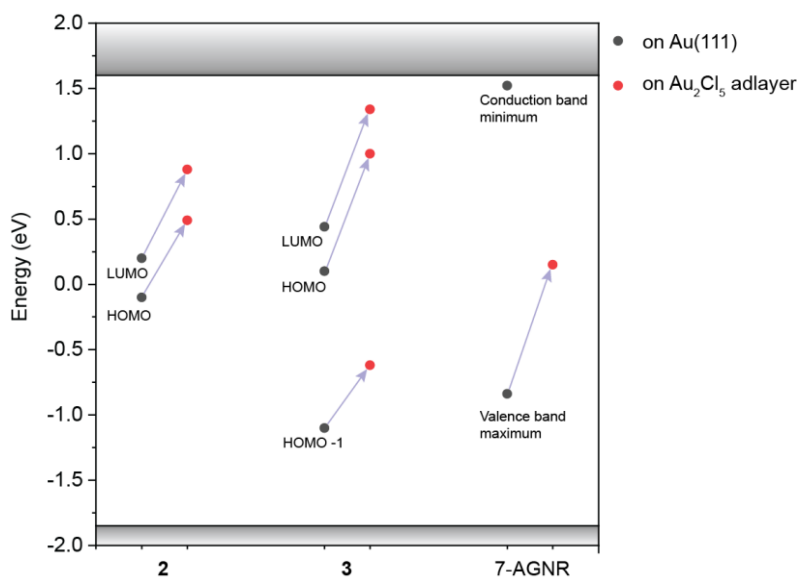

**Figure S2:** Graphical representation of the data shown in Table 1 in the main text. The energy at which the orbitals are observed on Au(111) (black circles) and upon $Au_2Cl_5$ adlayer intercalation (red circles) are shown for **2** and **3**. The 7-AGNR's valence band and conduction band extrema on Au(111) and the valence band maximum after intercalation are also shown for completeness. The band edges of the $Au_2Cl_5$ adlayer are indicated by shaded rectangles.

## S4: Particle-in-a-box states: Real-space comparison

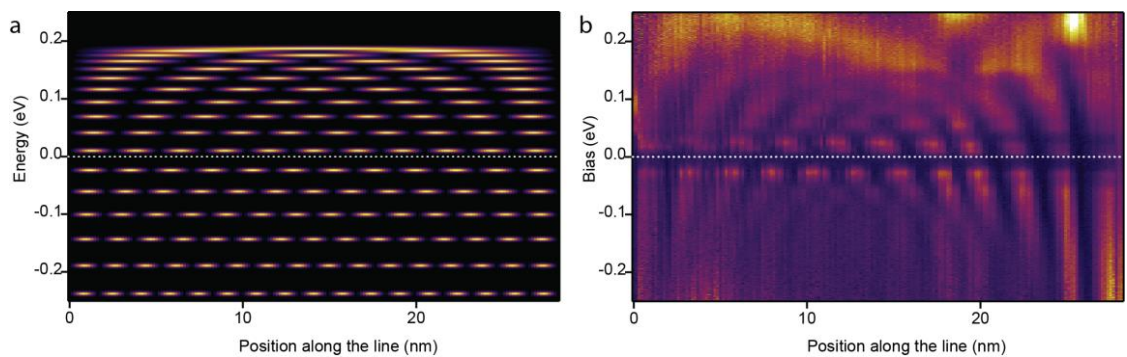

**Figure S3: a**, Probability density plot of particle-in-a-box states. The states are plotted based on the fitted parameters, i.e. box length = 28 nm, holes with an effective mass of 0.32 $m_e$ and band maximum of 0.187 eV. **b**, Experimental data from Fig. 4b.

## S5: Au$_2$Cl$_5$ adlayer intercalation occurs at room temperature

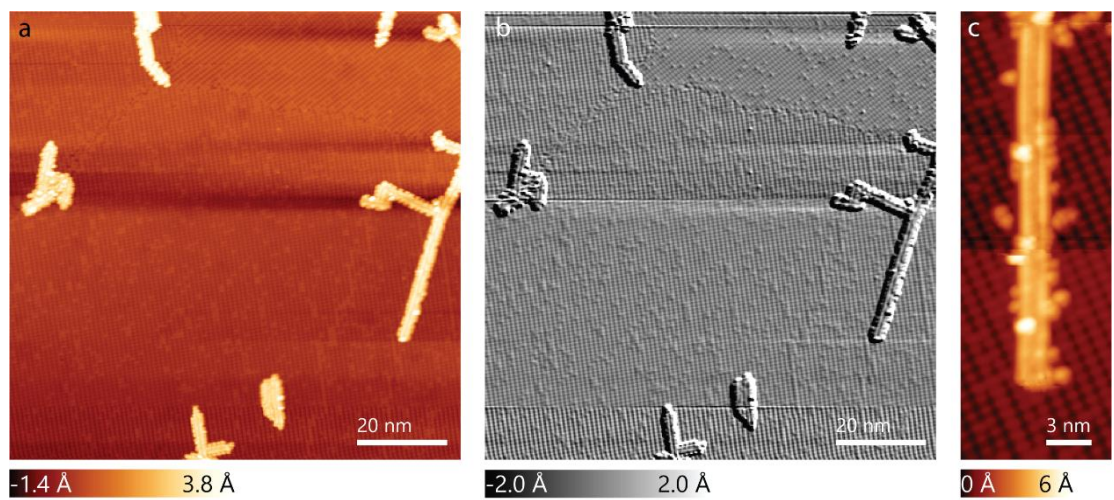

**Figure S4: Au$_2$Cl$_5$ adlayer intercalation at room temperature. a**, Overview STM image acquired on a 7-AGNR sample after Au$_2$Cl$_5$ adlayer intercalation at room temperature. The image shows the distinctive rows of the Au$_2$Cl$_5$ adlayer (V = -1.5 V, I = 20 pA). **b**, To better illustrate that the underlying Au$_2$Cl$_5$ adlayer structure is continuous under the GNRs, we numerically differentiate the image along the South-West direction. The gradient image shows three domains of the Au$_2$Cl$_5$ adlayer, and within a domain the rows of the Au$_2$Cl$_5$ adlayer are unaffected by the overlying GNRs. This shows that the Au$_2$Cl$_5$ adlayer has intercalated under the GNRs. **c**, STM image of a 7-AGNR with some physisorbed Cl atoms on the edges. Stable imaging conditions are prevented by the weakly adsorbed Cl (V = 2 V, I = 20 pA).

## S6: Formation of the Au$_2$Cl$_5$ adlayer and its intercalation underneath GNRs

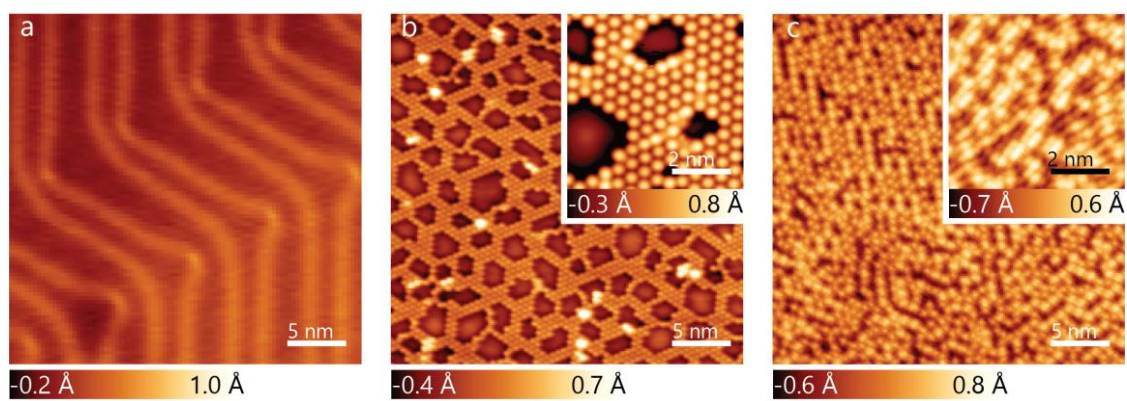

**Figure S5: Formation of the Au$_2$Cl$_5$ adlayer on Au(111). a**, STM image of the Au(111) surface exhibits the characteristic herringbone reconstruction (tunneling parameters: V = 2 V, I = 100 pA). **b**, After deposition of AuCl for 30 min, the STM image reveals the $\sqrt{3} \times \sqrt{3}$ superstructure of Cl chemisorbed on Au(111) (V = -1 V, I = 20 pA). Inset: higher resolution image of the superstructure (V = -0.5 V, I = 250 pA). **c**, STM image acquired after additional 30 min deposition of AuCl with the Au-adatoms incorporating into AuCl$_2$ molecules (V = -0.5 V, I = 10 pA). Inset: higher resolution image of the disordered structures (V = -0.5 V, I = 20 pA). An additional 30 min deposition of AuCl leads to the Au$_2$Cl$_5$ adlayer discussed in the main text (Fig. 2).

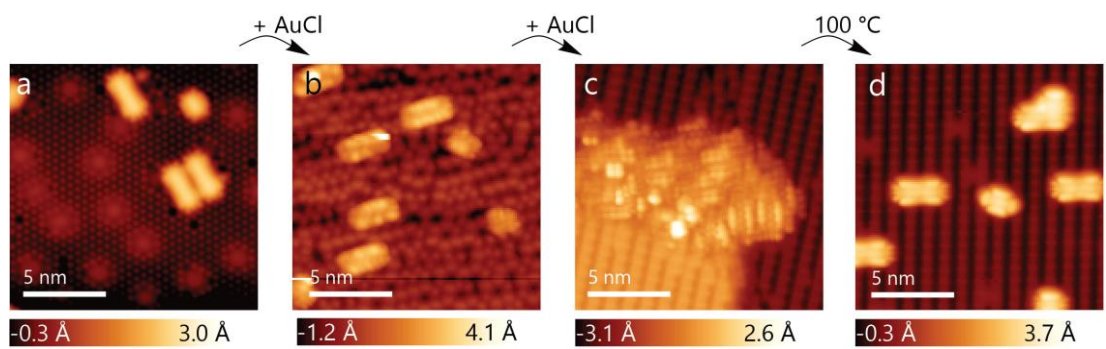

**Figure S6: Au$_2$Cl$_5$ adlayer intercalation under short 7-AGNR segments. a**, Short 7-AGNR segments (**2** and **3**, see main text) are synthesized on Au(111). The surface was allowed to cool down to room temperature and then AuCl sublimated onto it for 30 min. STM images acquired after the deposition of AuCl show an incomplete formation of the $\sqrt{3} \times \sqrt{3}$ superstructure as seen in Fig. S5. (V = 0.5 V, I = 50 pA). **b**, STM images acquired after depositing AuCl for an additional 30 min show the incorporation of Au atoms into a disordered gold chloride adlayer (V = 0.5 V, I = 20 pA). **c**, STM images acquired after an additional 30 min deposition of AuCl show the formation of the Au$_2$Cl$_5$ adlayer underneath the molecules. Excess Cl stabilizes self-assembled islands of **2** and **3** (V = 0.5 V, I = 50 pA). **d**, After annealing to 100 °C, excess Cl has desorbed and individual molecules of **2** and **3** are found on the Au$_2$Cl$_5$ layer (V = 2 V, I = 50 pA).